\documentclass[reprint,amsmath,amssymb,aps]{revtex4-2}

\usepackage{xcolor}
\usepackage{ulem}
\usepackage{lipsum}

\usepackage{graphicx}
\usepackage{dcolumn}
\usepackage{bm}

\newcommand{\be}{\begin{equation}} 
\newcommand{\ee}{\end{equation}}
\newcommand{\bea}{\begin{eqnarray}}
\newcommand{\eea}{\end{eqnarray}}

\usepackage{lineno,hyperref}
\modulolinenumbers[5]

\usepackage[english]{babel}

\usepackage[utf8]{inputenc}

\usepackage{soul}  

\newcommand{\bh}{{\bf{h}}}
\newcommand{\bht}{{\tilde {\bf{h}}}}
\newcommand{\bI}{{\bf{I}}}
\newcommand{\bc}{{\bf{c}}}
\newcommand{\bJ}{{\bf{J}}}
\newcommand{\bS}{{\bf{S}}}
\newcommand{\bQk}{\hat{\bf{Q}}}
\newcommand{\bHk}{\hat{\bf{H}}}
\newcommand{\bCk}{\hat{\bf{C}}}
\newcommand{\bB}{\tilde{\bf B}}
\newcommand{\bK}{{\bf K}}
\newcommand{\bb}{{\bf b}}
\newcommand{\bq}{{\bf q}}
\newcommand{\bone}{\bf 1}

\def\bal{\mbox {\boldmath ${\alpha}$}}
\def\betb{\mbox {\boldmath ${\beta}$}}

\begin{document} 
	
	
	\title{Analytic solution of the multidensity Ornstein-Zernike equation for
	hard-sphere fluid with tetrahedral quadrupolar-like surface adhesion}
	
	\author{Y. V. Kalyuzhnyi}
	\affiliation{University of Ljubljana, Faculty of Chemistry and Chemical Technology, Ve\v cna pot 113, Ljubljana, Slovenia,
		\\Yukhnovskii Institute for Condensed Matter Physics, Svientsitskoho 1, 79011 Lviv, Ukraine}
\thanks{Yuriy.Kalyuzhnyy@fkkt.uni-lj.si, yukal@icmp.lviv.ua }
	\author{P. T. Cummings}%
	\email{P.Cummings@hw.ac.uk}
	\affiliation{School of Engineering and Physical Sciences, Heriot-Watt University, Edinburgh EH14 4AS, UK}%
	
	
	
	\date{\today}
	
\begin{abstract}
		
We develop a multidensity formulation of the Ornstein–Zernike equation with Percus–Yevick closure for hard spheres with anisotropic surface adhesion of tetrahedral quadrupolar-like symmetry. An analytical solution is obtained using the invariant expansion method combined with Baxter’s factorization technique. Structural properties are evaluated using both the multidensity theory and the previously proposed single-density molecular OZ approach. At low stickiness, the two theories yield nearly identical predictions, while increasing stickiness leads to growing discrepancies and eventual loss of convergence of the single-density approach. These results highlight the importance of multidensity descriptions for strongly associating anisotropic fluids.
	
	\end{abstract}
	
	\maketitle

\section{Introduction.}

A common feature of the effective interaction between colloidal particles is the presence of a short-ranged attractive interaction, whose range is much smaller than the characteristic size of the particles. The minimal model capable of capturing this feature is the sticky hard-sphere (SHS) model of Baxter \cite{baxter1968percus}. Originally, the model was introduced to describe a fluid of hard spheres interacting via a relatively narrow square-well potential. In the SHS limit, this interaction is represented by a hard-sphere fluid supplemented by an additional sticky potential—an infinitely narrow and infinitely deep square-well acting only at contact. The limiting procedure is performed such that the second virial coefficient remains constant, allowing one to relate the SHS model to the corresponding square-well potential.
A key advantage of the Baxter model lies in its analytical tractability: it admits a closed analytical solution of the Ornstein–Zernike equation with the Percus–Yevick closure \cite{baxter1968percus}. This makes the SHS model a particularly convenient reference system for studying the phase behavior, structure, and thermodynamics of adhesive colloids.

More recently, various extensions of the original isotropic model—such as models with sticky point interactions or orientation-dependent sticky interactions—together with the corresponding generalizations of the analytical methods used to describe them, have been developed to account for the anisotropy present in real colloidal and molecular systems. The model with sticky points can be viewed as a limiting case of a hard-sphere model with additional off-center square-well sites whose wells are infinitely narrow and infinitely deep, acting at hard-sphere contact when the two particles adopt the appropriate mutual orientation. As in the Baxter SHS model, this limit is taken assuming that the second virial coefficient remains constant. The version of the model describing dimerization (one sticky point per particle) was proposed and analyzed by Wertheim \cite{wertheim1986fluids}, who obtained an analytical solution using the two-density PY approximation. Appropriate extensions of the original model and its analytical treatment within the multidensity PY approximation have also been proposed. These include multicomponent versions of the model, as well as models featuring multiple sticky sites per particle \cite{kalyuzhnyi1993solution,kalyuzhnyi1995solution,chang1995correlation,chang1995wertheim,vakarin1997integral}. 

Perhaps the first model incorporating orientationally dependent stickiness was introduced by Cummings and Blum \cite{cummings1986analytic}. They proposed a tetrahedral quadrupolar sticky hard-sphere model and, within the framework of the invariant expansion method, obtained an analytical solution of the molecular Ornstein–Zernike equation using the corresponding PY approximation. Blum et al. \cite{blum1990general} introduced a model of dipolar hard spheres with an additional dipolar-like sticky interaction and obtained an analytical solution within a mixed mean-spherical and Percus–Yevick (MSA/PY) approximation. A multicomponent version of this model was subsequently studied by Protsykevytch \cite{protsykevich2003multi}.
A further modification of the sticky dipolar-like hard-sphere model was later proposed by Gazzillo et al. 
\cite{gazzillo2008fluids,gazzillo2009local,gazzillo2010dipolar}, who extended its analytical description within the PY approximation to incorporate additional anisotropic features.

A common feature of the models discussed above is the presence of a short-ranged, strongly attractive interaction, which causes the particles to associate and form clusters of different types and geometries. This feature poses a significant challenge for their theoretical description. 
Traditional integral equation theories of liquids, which rely on a single-density description, are generally not well suited for treating strongly interacting systems \cite{kalyuzhnyi1991integral}. In the case of models with sticky points, this difficulty is circumvented by employing the multi-density integral equation formalism developed by Wertheim 
\cite{wertheim1984fluids1,wertheim1984fluids2,wertheim1986fluids1,wertheim1986fluids2}. However, models with orientationally dependent stickiness have so far been treated within the framework of traditional single-density integral equation approaches, which suggests that the applicability of the resulting theories is likely restricted to low and intermediate values of the stickiness. Unfortunately, the corresponding theoretical predictions have not yet been tested against exact computer simulation results. A straightforward application of Wertheim’s multidensity approach to these models is not possible, as this formalism explicitly relies on the presence of off-center square-well sites. Important feature of the theory is that it is formulated for the particles with a multiple number of square-well singly bondable sites, i.e. each site can form only one bond with the site of the other particle. 

In this paper, we consider the model with tetrahedral, quadrupolar-like surface adhesion proposed by Cummings and Blum \cite{cummings1986analytic}. To describe this system, we employ an alternative multidensity approach that accommodates orientationally dependent stickiness. This approach, developed by Kalyuzhnyi and Stell \cite{kalyuzhnyi1993effects}, can be viewed as a suitable reformulation of the multidensity formalism for particles possessing a single, multiply bondable square-well site.
We formulate the corresponding multidensity Ornstein–Zernike (OZ) equation with Percus–Yevick (PY) closure and obtain an analytical solution using the invariant expansion method combined with Baxter’s factorization technique. Numerical calculations are then performed using both our solution of the multidensity OZ equation and the previously developed solution of the single-density molecular OZ equation \cite{cummings1986analytic}. The resulting pair correlation functions obtained from the single- and multidensity approaches are compared and analyzed.

The paper is organized as follows. In Section II, we describe the interaction potential model, and in Section III we present the theoretical framework together with the solution of the multidensity OZ equation. Section IV contains the results and their discussion, while our conclusions are summarized in Section V.

\section{The model of hard spheres with anisotropic surface adhesion}

We consider the model represented by the fluid of hard spheres with additional
anisotropic square-well surface adhesion \cite{cummings1986analytic}. The interpartical pair potential of the model is
\be
U(12)=U_{hs}(r)+U^{(sw)}_{as}(12),
\label{U}
\ee
where $U_{hs}(r)$ is potential of hard spheres of the size $\sigma$ and 
$U^{(sw)}_{as}(12)$ is potential of square-well interaction, which is defined 
by its Mayer function
\be
f^{(sw)}_{as}(12)=\exp{\left[-\beta U^{(sw)}_{as}(12)\right]}-1={\sigma\over 12\tau}
{\Lambda(12)\over \Delta}\Theta(r,\sigma)
\label{Lambda}
\ee
where $\Theta(r,\sigma)=\theta(r-\sigma)\theta(\sigma+\Delta-r)$,
$\theta(x)$ is the Heaviside step function, $\Delta$ is the width of the square-well potential,
\be
\Lambda(12)=
\Phi_{00}^{000}(\omega_1\omega_2\omega_r)
+\lambda\sum_{\mu\nu}\Phi_{\mu\nu}^{224}(\omega_1\omega_2\omega_r),
\label{def}
\ee 
$r=|{\vec{\textbf{{r}}}}_{1}-{\vec{\textbf{{r}}}}_{2}|$ is the distance between centers of the two hard spheres, $1$ and $2$ denote position ($\vec{\textbf{{r}}}_1$ and $\vec{\textbf{{r}}}_2$) and 
orientation  ($\omega_1$ and $\omega_2$) of the particles 1 and 2,
$\omega_r$ denotes for the orientation of the vector joining their centers,
$\Phi_{\mu\nu}^{mnl}$ is rotational invariant, $\delta(\ldots)$ is 
Dirac delta-function, $\lambda$ is a parameters specifying the relative 
strength of spherically symmetric and orientationally dependent parts of the 
sticky interaction, $\tau$ is dimensionless measure of the temperature and
summation in (\ref{def}) is carried out for 
$\mu\nu=22,2\emph{2},\emph{2}2,\emph{2}\emph2$ ($\emph{2}=-2$).
Here $0\leq\lambda\leq 1$,  
$\omega_1$, $\omega_2$ and $\omega_r$ represent each the set of three 
Euler angles. 

To enable an analytical treatment of the model, we approximate the Mayer 
function $f_{as}^{(sw)}(12)$ by a Dirac delta function while keeping the
value of the second virial coefficient constant. This is achieved by 
considering the limiting process similar to that suggested by Baxter \cite{baxter1968percus}.
We have
\be
f_{as}^{(sw)}(12)\approx f_{as}^{(st)}(12)=
{\sigma\over 12\tau}\Lambda(12)\delta(r-\sigma),
\label{approxdelta}
\ee
where $f_{as}^{(st)}(12)$ is the Mayer function corresponding to the sticky 
potential and $\delta(x)$ denotes the Dirac delta function.

\be
\beta U^{sw}_{as}(12)=-\theta(r-\sigma)\theta(\sigma+\Delta-r)
\ln{\left[1+{\Lambda(12)\over 12\tau}{\sigma\over\Delta}\right]},
\ee

According to the expression for $\Lambda(12)$ (\ref{def}) the symmetry of the model potential (\ref{U}) is the same as that of the 
quadrupolar part of the pair potential between particles with two positive 
and two negative  charges located at the corners of the regular tetrahedron. 
In particular at fixed $\omega_r$ there are eight sets of Euler angles 
$\omega^{(i)}_1\omega^{(i)}_2\;(i=1\ldots 8)$, with the values that 
specify configurations when $\Lambda(12)$ reaches its maximum value of
$1+\lambda$. Thus our model can be seen as a fluid of hard spheres
with two "negative" and two "positive" sticky spots arranged tetrahedrally
on the surface of each particle. The sticky interaction is valid only between the 
spots of different type.

\section{Theory}

\subsection{APY closure relations}

The thermodynamic and structural properties of the present model are analyzed within the framework of the associative Percus–Yevick (APY) approximation combined with the so-called ideal network approximation 
\cite{kalyuzhnyi1993effects,kalyuzhnyi2023integral}. The theoretical description is based on the multidensity Ornstein–Zernike (OZ) equation,
\be
{\bh} (12)=\bc(12)+{\rho\over 8\pi^2}\int \bc(13)\bal\bh(32)
d{\vec{\textbf{{r}}}}_3d\omega_3,
\label{WOZ}
\ee
together with the APY closure relation,
\be
\left\{\begin{array}{ccc}
	\bh(12)=-{\bf E}+
	{\bf B}(12)
	\delta(r-\sigma^-),\;\; &r&<\sigma \\
	\bc(12)=\;\;\;0,
	\hspace{25mm}\;\;\;\;\;\; &r&>\sigma
\end{array} \right.,
\label{conical}
\ee
where 
${\bh}(12)$, ${\bc}(12)$, ${\bal}$, ${\bf E}$ and ${\bf B}$(12) 
denote the following matrices:
\be
\bh=
\begin{pmatrix}
	h_{00}  & h_{01} \\
	h_{10}  & h_{11} \\
\end{pmatrix},
\;\;\;
\bc=
\begin{pmatrix}
	c_{00}  & c_{01} \\
	c_{10}  & c_{11} \\
\end{pmatrix},
\ee
\be
\bal=
\begin{pmatrix}
	\alpha_{00} & \alpha_{01} \\
	\alpha_{10} &    \alpha_{11} \\
\end{pmatrix}
=
\begin{pmatrix}
	1 & \sum_{i=0}^{n_s-1}x_i \\
	\sum_{i=0}^{n_s-1}x_i &  (1-\delta_{1n_s})\sum_{i=0}^{n_s-2}x_i \\
\end{pmatrix},
\label{matrix}
\ee
\be
{\bf E}=
\begin{pmatrix}
	1 & 0 \\
	0 &    0 \\
\end{pmatrix},\;\;\;
{\bf B}(12)=
\begin{pmatrix}
	0 & 0 \\
	0 & \Lambda(12)\sigma g_{00}(\sigma^+) \\
\end{pmatrix},
\label{EB}
\ee
$x_i$ represents the fraction of particles bonded $i$ times and determined from the 
set of equations
\be
\left\{
\begin{array}{lcc}
	x_1=2\eta a x_0g_{00}(\sigma^+)\left(1-{1\over n_s!}{x_1^{n_s}\over x_0^{n_s-1}}\right)/\tau
	\\
	x_0^{n_s-1}=\sum_{i=0}^{n_s}{x_0^ix_1^{n_s-i}\over (n_s-i)!}
\end{array}
\right.,
\label{mal}
\ee
$h_{ij}(12)$ and $c_{ij}(12)$ denote, respectively, the partial total and 
direct correlation functions, $g_{ij}(12)=h_{ij}(12)+\delta_{i0}\delta_{0j}$ 
and denotes partial pair distribution function. Here the lower indices $i$ 
and $j$ take the values 0 and 1 and denote bonding state of the 
corresponding particle, i.e. 0 denotes an unbonded state and 1 denotes 
singly bonded state. Note that the partial correlation functions between 
unbonded particles $h_{00}(r)$ and $c_{00}(r)$ depend only on the mutual distance $r$. For the pair correlation function $h(12)$ we have
\be
h(12)=\sum_{ij=0}^1 \alpha_{0i}h_{ij}(12)\alpha_{j0}.
\label{totalh}
\ee
Multidensity OZ equation (\ref{WOZ}), APY closure relation (\ref{conical}) and relation 
between densities (\ref{mal}) form a closed set of equations to be solved.

To solve the above set of equations, we employ the Baxter factorization method 
\cite{baxter1970ornstein} in combination with the rotationally invariant expansion technique \cite{blum1972invariant,blum1972invariant2,blum1973invariant}, which enables a systematic treatment of the angular dependence of the correlation functions. 
Following earlier study \cite{cummings1986analytic}, we assume that the only nonzero harmonics in the correlation functions $\bh(12)$ and $\bc(12)$
are those appearing in the expression for $f^{(sw)}_{as}(12)$
[Eq. (\ref{approxdelta})], as well as those that can be generated from this initial set through successive angular convolutions. Accordingly, our analysis is limited to the set of rotationally invariant laboratory frame \cite{gray2011theory} 
expansion coefficients ${\bf f}^{mnl}_{\mu\nu}(r)$ with the following values 
of their indices:
 $(mnl\mu\nu)=(00000),(22l22),(22l2\emph{2}),(22l\emph{2}2),(22l\emph{2}\emph{2})$,
 where ${\bf f}\equiv {\bf c},{\bf h}$ and $l=0,2,4$. In terms of these coefficients  APY closure relation (\ref{conical}) is
\be
\left\{\begin{array}{cc}
	\bh_{00}^{000}(r)=-{\bf E}+{\bf B}_{00}^{000}\delta(r-\sigma^-),\; r<&\sigma \\
	\bh_{\mu\nu}^{224}(r)={\bf B}_{\mu\nu}^{224}\delta(r-\sigma^-),\hspace{10mm} r<&\sigma \\
	\bh^{22l}_{\mu\nu}(r)=0,\;\;\;\;\;(l=0,2),\hspace{9mm}\; r<&\sigma \\
	\bc_{\mu\nu}^{22l}(r)=0,\;\;(l=0,2,4),\hspace{9mm}\; r>&\sigma
\end{array} 
\right.,
\label{conical1}
\ee
where
$\bh_{\mu\nu}^{mnl}(r)$, $\bc_{\mu\nu}^{mnl}(r)$ and ${\bf B}_{\mu\nu}^{mnl}$ denote matrices with the elements
\[
[{\bf h}^{mnl}_{\mu\nu}(r)]_{ij}= h^{mnl}_{\mu\nu;ij}(r),\;
[{\bf c}^{mnl}_{\mu\nu}(r)]_{ij}= c^{mnl}_{\mu\nu;ij}(r)
\]
\[
[{\bf B}_{00}^{000}]_{ij}=B_{00,ij}^{000}=
(1-\delta_{i0})(1-\delta_{j0})\sigma g_{00}(\sigma^+)/(12\tau), 
\]
\[[{\bf B}_{\mu\nu}^{224}]_{ij}=B_{\mu\nu,ij}^{224}=\lambda(1-\delta_{i0})(1-\delta_{j0})\sigma g_{00}(\sigma^+)/(12\tau),
\] 
$i=0.1$, $j=0,1$ and
$\mu\nu=22,2\emph{2},\emph{2}2,\emph{2}\emph2$.
Taking into account the symmetry of the model we have:
\be
{\bf f}^{22l}_{22}(r)={\bf f}^{22l}_{2\emph{2}}(r)=
{\bf f}^{22l}_{\emph{2}2}(r)={\bf f}^{22l}_{\emph{2}\emph{2}}(r),
\label{symm}
\ee
where ${\bf f}\equiv {\bf c},{\bf h}$ and $l=0,2,4$.
Thus in what follows we will be focused on the calculation of the 
expansion coefficients ${\bf h}^{000}_{00}(r)$ 
 and ${\bf h}^{22l}_{22}(r)$
(${\bf c}^{000}_{00}(r)$ and ${\bf c}^{22l}_{22}(r)$).
Corresponding versions of the multidensity OZ equation (\ref{WOZ}) 
written in Fourier $k$-space in terms of the intermolecular 
$k$-frame harmonic expansion coefficients are
\be
\left[{\bf 1}+2\rho{\hat {\bf J}}^{00}_{00;0}(k)\bal\right]
\left[(\bal)^{-1}-2\rho{\hat {\bf S}}^{00}_{00;0}(k)\right]=
(\bal)^{-1},
\label{conical2}
\ee
\be
\left[{\bf 1}+4\rho{\hat {\bf J}}^{22}_{22;\chi}(k)\bal\right]
\left[(\bal)^{-1}-4\rho{\hat {\bf S}}^{22}_{22;\chi}(k)\right]=
(\bal)^{-1},
\label{conical3}
\ee
where 
\be
{\hat {\bf F}}^{mn}_{\mu\nu;\chi}(k)=
{1\over 2}
\int_0^\infty dr\;\left[e^{ikr}{\bf F}^{mn}_{\mu\nu;\chi}(r)+e^{-ikr}
{\bf F}^{nm}_{\nu\mu;\chi}(r)\right]
\label{Jk}
\ee
\be
{\bf F}^{mn}_{\mu\nu;\chi}(r)=2\pi (-)^\chi\sum_{l}
\begin{pmatrix}
	m & n & l \\ 
 \chi &  \underline{\chi} &0 \\
\end{pmatrix}
 \int_r^\infty dt\;t P_l(r/t){\bf f}_{\mu\nu}^{mnl}(t),
\label{J}
\ee
$({\bf F},{\bf f})=({\bf J},{\bf h})$ or 
$({\bf F},{\bf f})=({\bf S},{\bf c})$, and
$P_l(x)$ is Legendre polynomial of the order $l$. 
In the subsequent analysis, we will make use of the “inverse’’ form of this 
equation, given by \cite{blum1973invariant}
\[
{\bf f}_{\mu\nu}^{mnl}(r)=-{2l+1\over 2\pi}\sum_{\chi=-\inf(m,n)}^{-\inf(m,n)}
(-)^\chi
\begin{pmatrix}
	m & n & l \\ 
	\chi &  \underline{\chi} &0 \\
\end{pmatrix}
\]
\[
\times\int_0^\infty dt\;{\bf F}^{mn}_{\mu\nu;\chi}(t)\left[{1\over r}\delta^{'}(r-t)
-{1\over r^2}P^{'}_l(1)\delta(r-t)\right.
\]
\be
\left.
+{1\over r^3}P^{''}_l(t/r)\theta(r-t)\right].
\label{inverse}
\ee
Equation {\ref{conical2})
and closure relations for the expansion coefficients ${\bf h}^{000}_{00}(r)$
and ${\bf c}^{000}_{00}(r)$ (\ref{conical1}) are identical to those solved in 
our previous paper \cite{kalyuzhnyi2023integral}. Therefore, we omit a detailed 
discussion here and direct the readers to the original publication. Next
we will be focused on the solution of the equation (\ref{conical3}).
In the ensuing analysis we will assume that all distances are measured in units 
of $\sigma$, i.e. $\sigma=1$. 

According to the closure relations (\ref{conical1}) all expansion coefficients
${\bf c}^{22l}_{\mu\nu}(r)$ are short-ranged, thus solution of equation
(\ref{conical3}) can be obtained using Baxter factorization method 
\cite{baxter1970ornstein}. Following Baxter, we have
\be
\left\{
\begin{array}{cc}
	(\bal)^{-1}-4\rho\hat{\bf S}^{22}_{22;\chi}(k)=4\rho^2\bQk_\chi(k)\bal\bQk^{T}_\chi(-k)\\
	4\rho^2\left[{\bf 1}+4\rho\hat{\bf J}^{22}_{22,\chi}(k)   
	\bal\right]\bQk_\chi(k)\bal\bQk_\chi^{T}(-k)=(\bal)^{-1}
\end{array}
\right.,
\label{fact}
\ee
where the superscript $T$ denotes matrix transpose, $\bQk_\chi(k)$ is Baxter factor function,
\be
2\rho\bQk_\chi(k)=\left(\bal\right)^{-1}-2\rho\int_0^1 {\bf Q}_\chi(r)e^{ikr}dr
\label{Qk} 
\ee
and the real space factor function ${\bf Q}_\chi(r)=0$ for $r>1$. Upon inverting 
of these equations to real space we have
\be
\left\{
\begin{array}{cc}
\bS^{22}_{22;\chi}(r)={\bf Q}_\chi(r)-2\rho\int_0^1{\bf Q}_\chi(t)\bal{\bf Q}_\chi(t-r)dt\\
\bJ^{22}_{22;\chi}(r)={\bf Q}_\chi(r)+2\rho\int_0^1\bJ^{22}_{22;\chi}(|r-t|)\bal{\bf Q}_\chi(t)dt
\end{array}
\right..
\label{rreal}
\ee
Taking into account the closure relations (\ref{conical1}) and using the expression for 
${\bf J}^{22}_{22;\chi}(r)$ (\ref{J}) we obtain
\be
{\bf J}^{22}_{22;\chi}(r)=\betb_{\chi,0}+\betb_{\chi,2}r^2+\betb_{\chi,4}r^4
-\bB_\chi\theta(r-1^-),
\label{Jcap}
\ee
where $0\leq r\leq 1$, 
\[
\betb_{\chi,0}=
(-)^\chi\left[\begin{pmatrix}
	2 & 2 & 0 \\ 
	\chi &  \underline{\chi} &0 \\
\end{pmatrix}
\bb^{220}_{22,0}\right.
\]
\be
\left.
-{1\over 2}
\begin{pmatrix}
	2 & 2 & 2 \\ 
	\chi &  \underline{\chi} &0 \\
\end{pmatrix}
\bb^{222}_{22,0}
+{3\over 8}
\begin{pmatrix}
	2 & 2 & 4 \\ 
	\chi &  \underline{\chi} &0 \\
\end{pmatrix}
\bb^{224}_{22,0}\right],
\ee
\be
\betb_{\chi,2}=(-)^\chi\left[
{3\over 2}
\begin{pmatrix}
	2 & 2 & 2 \\ 
	\chi &  \underline{\chi} &0 \\
\end{pmatrix}
\bb^{222}_{22,2}
-{30\over 8}
\begin{pmatrix}
	2 & 2 & 4 \\ 
	\chi &  \underline{\chi} &0 \\
\end{pmatrix}
\bb^{224}_{22,2}\right],
\ee
\be
\betb_{\chi,4}=(-)^\chi
{35\over 8}
\begin{pmatrix}
	2 & 2 & 4 \\ 
	\chi &  \underline{\chi} &0 \\
\end{pmatrix}
\bb^{224}_{22,4},
\ee
\be
\bb^{mnl}_{\mu\nu,p}=2\pi\int_0^\infty dt\;t^{1-p}\bh^{mnl}_{\mu\nu}(t),
\label{bs}
\ee
\be
\bB_\chi=2\pi(-)^\chi
\begin{pmatrix}
	2 & 2 & 4 \\ 
	\chi &  \underline{\chi} &0 \\
\end{pmatrix}
{\bf B}^{224}_{22}.
\label{Btilde}
\ee
Substituting expression for $\bJ^{22}_{22;\chi}$ into second of equations (\ref{rreal}) 
yields
\be
{\bf Q}_\chi(r)=\sum_{m=1}^4{1\over m!}\bq_{\chi,m}(r^m-1)+\bB_\chi[1-\theta(r-1^-)]
\label{Qc}
\ee
The coefficients $\bq_{\chi,m}$, which appear in the expression for ${\bf Q}_\chi(r)$ 
(\ref{Qc}) satisfy the following set of linear equations
\be
\left\{
\begin{array}{cc}
\bq_{\chi,1}-4\rho\betb_{\chi,2}\bal\bK_{\chi,1}-8\rho\betb_{\chi,4}\bK_{\chi,3}=0\\
{1\over 2}\bq_{\chi,2}-\betb_{\chi,2}+2\rho\betb_{\chi,2}\bal\bK_{\chi,0}+12\rho\betb_{\chi,4}\bal\bK_{\chi,2}=0\\
{1\over 6}\bq_{\chi,3}-8\rho\betb_{\chi,4}\bal\bK_{\chi,1}=0\\
\betb_{\chi,4}-{1\over 24}\bq_{\chi,4}-2\rho\betb_{\chi,4}\bal\bK_{\chi,0}=0
\end{array}
\right.,
\label{linear}
\ee
obtained upon substitution of the expression for ${\bf Q}_{\chi}(r)$ (\ref{Qc}) into
the second of the equations (\ref{rreal}) and using the expression for 
${\bf J}_{22;\chi}^{22}(r)$ (\ref{Jcap}). Here $\bK_{\chi,n}$ is the $n$th moment of 
the factor function ${\bf Q}_\chi(r)$
\be
\bK_{\chi,n}=\int_0^1 dt\;t^n{\bf Q}_\chi(r),
\label{moments}
\ee
i.e.
\be
\begin{pmatrix}
	-\bK_{\chi,0} \\ 
	-2\bK_{\chi,1} \\
	-3\bK_{\chi,2} \\
	-4\bK_{\chi,3} \\
\end{pmatrix}
=
\begin{pmatrix}
	1/2&1/3&1/4&1/5& \\ 
	1/3&1/4&1/5&1/6& \\ 
	1/4&1/5&1/6&1/7& \\ 
	1/5&1/6&1/7&1/8& \\ 
\end{pmatrix}
\begin{pmatrix}
	\bq_{\chi,1} \\ 
	\bq_{\chi,2} \\
	{1\over 2}\bq_{\chi,3} \\
	{1\over 6}\bq_{\chi,4} \\
\end{pmatrix}
-
\begin{pmatrix}
	\bB_{\chi} \\ 
	\bB_{\chi} \\
	\bB_{\chi} \\
	\bB_{\chi} \\
\end{pmatrix}.
\ee
The solution to the system of equations (\ref{linear}) can be written in the 
following form:
\be
\begin{pmatrix}
	\bq_{\chi,1} \\ 
	\bq_{\chi,2} \\
	\bq_{\chi,3} \\
	\bq_{\chi,4} \\
\end{pmatrix}
={\bf M}^{-1}_\chi
\begin{pmatrix}
	2\rho(\betb_{\chi,2}+\betb_{\chi,4})\bal\bB_\chi \\ 
	\betb_{\chi,2}-2\rho(\betb_{\chi,2}+2\betb_{\chi,4})\bal\bB_\chi \\
	4\rho\betb_{\chi,4}\bal\bB_\chi \\
	\betb_{\chi,4}-2\rho\betb_{\chi,4}\bal\bB_\chi \\
\end{pmatrix},
\label{solution}
\ee

\onecolumngrid
\noindent\rule{0.5\linewidth}{0.4pt}\\
where
\be
{\bf M}_\chi=
\begin{pmatrix}
\bone+\rho({2\over 3}\betb_{\chi,2}+{2\over 5}\betb_{\chi,4})\bal&
\rho({1\over 2}\betb_{\chi,2}+{1\over 3}\betb_{\chi,4})\bal&
\rho({1\over 5}\betb_{\chi,2}+{1\over 7}\betb_{\chi,4})\bal&
\rho({1\over 18}\betb_{\chi,2}+{1\over 24}\betb_{\chi,4})\bal&\\
-\rho(\betb_{\chi,2}-\betb_{\chi,4})\bal&
{1\over 2}\bone-\rho({2\over 3}\betb_{\chi,2}+{4\over 5}\betb_{\chi,4})\bal&
-\rho({1\over 4}\betb+{1\over 3}\betb_{\chi,4})\bal&
-\rho({1\over 15}\betb_{\chi,2}+{2\over 21}\betb_{\chi,4})\bal\\
{4\over 3}\rho\betb_{\chi,4}\bal&
\rho\betb_{\chi,4}\bal&
{1\over 6}\bone+{2\over 5}\rho\betb_{\chi,4}\bal&
{1\over 9}\rho\betb_{\chi,4}\bal\\
-\rho\betb_{\chi,4}\bal&
-{2\over 3}\rho\betb_{\chi,4}\bal&
-{1\over 4}\rho\betb_{\chi,4}\bal&
{1\over 24}\bone-{1\over 15}\rho\betb_{\chi,4}\bal\\
\end{pmatrix}
\ee
\noindent
\hfill\rule{0.5\linewidth}{0.4pt}\\
\twocolumngrid
\noindent
and $[\bone]_{ij}=\delta_{ij}$
This solution represent coefficients $\bq_{\chi,m}$ in terms of six parametetrs 
$\betb_{\chi,2}$ and $\betb_{\chi,4}$, which in turn are the 
functions of three unknown integrals $\bb_{22,2}^{222}$, $\bb_{22,2}^{224}$ and 
$\bb_{22,4}^{224}$. From the closure relations (\ref{conical1}) it follows that the harmonic expansion coefficients of the direct correlation function vanish for 
$r>1$, i.e.
\be
\left\{
\begin{array}{cc}
\bc_{22}^{222}(r)=0,\;\;\;\;{\rm r}>1\\
\bc_{22}^{224}(r)=0,\;\;\;\;{\rm r}>1
\end{array}
\right.
\label{closurec}
\ee

Taking into account 
these conditions and 
using  equation 
(\ref{inverse}) to calculate $\bc_{22}^{222}(r)$ and
$\bc_{22}^{224}(r)$ for $r>1$ 
we get the 
set of the
following three equations
\be
\left\{
\begin{array}{cc}
    \bI_{0,0}+\bI_{1,0}-2\bI_{2,0}=0\\
    \bI_{0,2}-{4\over 3}\bI_{1,2}+{1\over 3}\bI_{2,2}=0\\
    \bI_{0,0}-{4\over 3}\bI_{1,0}+{1\over 3}\bI_{2,0}=0
\end{array},
\right.
\label{3b}
\ee
where
$\bI_{\chi,m}=\int_0^1 dt\;t^m\bS_{22;\chi}^{22}(t)$. These integrals can be 
calculated using first of the equations (\ref{rreal}) and expression for 
${\bf Q}_\chi(r)$ (\ref{Qc}), we have

\onecolumngrid
\noindent
\rule{0.5\linewidth}{0.4pt}\\
\[
\bI_{\chi,0}=-\sum_{m=1}^4{\bq_{\chi,m}\over (m-1)!(m+1)}
+2\rho\sum_{mn=1}^4{\bq_{\chi,m}\bal\bq_{\chi,n}^T\over n!m!}
\left[{m\over (n+1)(n+2)(m+n+2)}
+{1\over m+2}-{1\over 2}\right]
\]
\be
+\rho\sum_m^4{\bq_{\chi,m}\bal\bB_\chi^T\over (m-1)!(m+2)}
-2\rho\bB_\chi\bal\sum_{m=1}^4{\bq^T_{\chi,m}\over m!}\left[{1\over (m+1)(m+2)}-{1\over 2}\right]-\rho\bB_\chi\bal\bB^T_\chi+\bB_\chi,
\label{I0}
\ee

\[
\bI_{\chi,2}=-\sum_{m=1}^4{\bq_{\chi,m}\over 3(m-1)!(m+3)}
+2\rho\sum_{mn=1}^4{\bq_{\chi,m}\bal\bq^T_{\chi,n}\over m!n!}
\left[{2m\over (n+1)(n+2)(n+3)(n+4)(m+n+4)}+{1\over 3(m+4)}-{1\over 12}\right]
\]
\be
+{1\over 6}\rho\sum_m^4{\bq_{\chi,m}\bal\bB_\chi^T\over (m-1)!}
-2\rho\bB_\chi\bal\sum_{m=1}^4{\bq^T_{\chi,m}\over m!}\left[{2\over (m+1)(m+2)(m+3)(m+4)}-{1\over 12}\right]
-{1\over 6}\rho\bB_\chi\bal\bB^T_\chi+{1\over 3}\bB_\chi.
\label{I2}
\ee
\noindent
\hfill\rule{0.5\linewidth}{0.4pt}
\twocolumngrid
\noindent

Thus, the solution of the APY approximation for the model under consideration
reduces to solving the set of equations (\ref{3b}) for
$\bb_{22,2}^{222}$, $\bb_{22,2}^{224}$, and $\bb_{22,4}^{224}$, provided that the 
fractions $x_i$, which enter the matrix $\bal$ are known. These fractions follow
from the solution of the set of equations (\ref{mal}), where the contact value of
the radial distribution function $g_{00}(r)$ is obtained from the solution of
the OZ equation (\ref{conical2}), derived earlier \cite{kalyuzhnyi2023integral}, i.e.
\be
g_{00}(1^+)=(1+\eta/2)/ (1-\eta)^2.
\ee

\subsection{Structure and thermodynamics}

The structure properties of the model under consideration are described by the 
harmonic expansion coefficients of the correlation function $\bh(12)$, i.e 
$\bh^{mnl}_{\mu\nu}(r)$. These coefficients can be obtained using the 
expression for the Baxter factor $Q$-function Eq. (\ref{Qc}), together with 
relations (\ref{fact}). In turn, these relations can be
written in terms of 
the intermolecular $k$-frame harmonic expansion 
coefficients for direct and total 
correlation functions, 
$\bHk^{mn}_{\mu\nu;\chi}(k)$ 
and
$\bCk^{mn}_{\mu\nu;\chi}(k)$, respectively.
 According to (\ref{Jk}) and (\ref{J}) 
\be
\bHk^{mm}_{\mu\mu;\chi}(k)=2{\hat \bJ}^{mm}_{\mu\mu;\chi}(k),\;\;
\bCk^{mm}_{\mu\mu;\chi}(k)=2{\hat \bS}^{mm}_{\mu\mu;\chi}(k)
\label{HS}
\ee
and we have 
\be
\left\{
\begin{array}{cc}
	(\bal)^{-1}-2\rho\hat{\bf C}^{22}_{22;\chi}(k)=4\rho^2\bQk_\chi(k)\bal\bQk^{T}_\chi(-k)\\
	4\rho^2\left[{\bf 1}+2\rho\hat{\bf H}^{22}_{22,\chi}(k)  
	\bal\right]\bQk_\chi(k)\bal\bQk_\chi^{T}(-k)=(\bal)^{-1}
\end{array}
\right.,
\label{facthc}
\ee
Using these relations we obtain 
\be
\bht^{22l}_{22}(r)
={(-i)^l\over 2\pi^2}\int_0^\infty k^2j_l(kr)
{\hat {\bf t}}_{22}^{22l}(k)dk,
\label{Hankel}
\ee

where $\bht^{22l}_{22}(r)$ is the regular part of $\bh^{22l}_{22}(r)$ (without delta-function term),
$r>1$, $j_l(x)$ is the $l$th order spherical Bessel function, 
\be
{\hat {\bf t}}_{22}^{22l}(k)=(2l+1)\sum_{\chi=-2}^{2}(-)^\chi
\begin{pmatrix}
	2&2&l& \\
	\chi & \underline{\chi} & 0& \\
\end{pmatrix}
{\hat {\bf T}}^{22}_{22;\chi}(k)
\label{T}
\ee
and
\[
{\hat {\bf T}}^{22}_{22;\chi}(k)={\hat {\bf H}}^{22}_{22;\chi}(k)
-{\hat {\bf C}}^{22}_{22;\chi}(k)
=-{1\over\rho}\bal^{-1}
\]
\be
+2\rho\bQk_{\chi}(k)\bal\bQk^T_{\chi}(-k)+
{1\over 8\rho^3}\left[\bal\bQk_{\chi}(k)\bal\bQk^T_{\chi}(-k)\bal\right]^{-1}.
\ee
\begin{figure}[htb!]
	\begin{center}
		\includegraphics[height=2.9cm,clip]{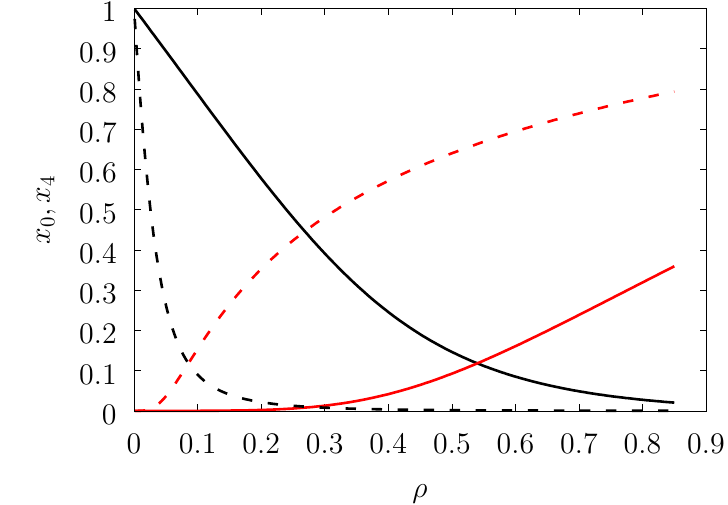}
		\includegraphics[height=2.9cm,clip]{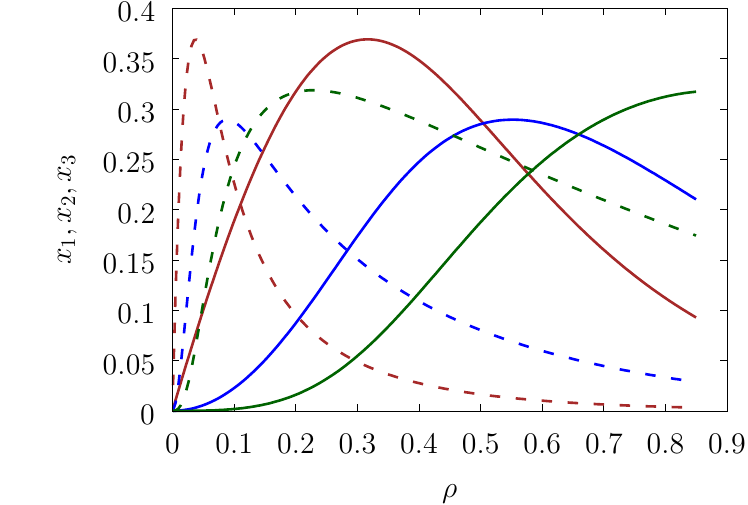}
	\end{center}
	\caption{Fractions of $i-$times bonded particles $x_i$: $x_0$ (black lines),
		$x_1$ (brown lines), $x_2$ (blue lines), $x_3$ (green lines) and
		$x_4$ (red lines) at $\tau=0.5$ (solid lines) and	$\tau=0.04$ (dashed lines).}
	\label{f1}
\end{figure}

The relation between harmonic expansion coefficients $h^{mnl}_{\mu\nu}(r)$ for the pair correlation function $h(12)$ and harmonic expansion coefficients for the partial
correlation functions, which are represented by the matrix $\bh^{mnl}_{\mu\nu}(r)$, is
\be
h^{mnl}_{\mu\nu}(r)=\sum_{ij=0}^1 \alpha_{0i}h^{mnl}_{\mu\nu;ij}(r)\alpha_{j0}.
\label{harmonictotalh}
\ee

The corresponding expression for harmonic expansion coefficients of the version of the
model with square-well potential can be obtained substituting delta function term in 
the closure relations (\ref{conical1}) by its finite counterpart, which gives
\be
\bh^{22l(sw)}_{22}(r)=\bht^{22l}_{22}(r)+{\bf B}_{22}^{22l}\Theta(r,1)/\Delta.
\label{soft}
\ee
Analytic expressions for $\bQk_{\chi}(k)$ can be easily derived using equations (\ref{Qk}) and (\ref{Qc}). Hankel transformation (\ref{Hankel}) have to be performed 
numerically \cite{talman1978numerical}. 

The thermodynamic properties of the model are evaluated using the energy route, 
following the scheme developed earlier \cite{kalyuzhnyi2020integral,kalyuzhnyi2015theoretical}. We begin with the general 
expression for the excess internal energy $\Delta E$, namely
\[
\beta{\Delta E\over N}=-{1\over 2\rho}\beta\int{\rho(12)\over e_{as}^{sw}(12)}
{\partial e_{as}^{sw}(12)\over\partial\beta}\;d{\vec{\textbf{{r}}}}_{12} 
d\omega_1d\omega_2=
\]
\be
-{1\over 2}\beta\rho\int e_{hs}(r)y(12){\partial e_{as}^{sw}(12)\over\partial\beta}
\;d{\vec{\textbf{{r}}}}_{12}d\omega_1d\omega_2
\label{energy1}
\ee
and concidering the pair density $\rho(12)$ as a sum of fugacity graphs, performing 
their resummation and analysis in terms of partial cavity correlation function
$y_{ij}(12)$. Here $e_{hs}(r)=\exp{[-\beta U_{hs}(r)]}$ and 
$e_{as}^{(sw)}(12)=\exp{[-\beta U_{as}^{(sw)}(12)]}$. The final expression for the
excess internal energy $\Delta E$ is
\be
\beta{\Delta E\over N}=-\alpha_{01}{x_1\over 2x_0}\left(1+12\Delta\tau\right)
\ln{\left(1+{1\over 12\Delta\tau}\right)}.
\label{energy2}
\ee
All the rest of thermodynamic properties are calculated using standard thermodynamic relations using expression for the internal energy (29) as an input.

\section{Results and discussion}

The system of nonlinear algebraic equations (\ref{3b}) was solved numerically using the standard Newton method. Note that the unknowns in this system are $2\times 2$
matrices; therefore, Eq. (\ref{3b}) represents a set of twelve coupled equations for the matrix elements. In the zero-density limit, the solutions reduce to
\be
[\bb^{222}_{22,2}]_{ij}=0,\;\;
[\bb_{22,2}^{224}]_{ij}=[\bb_{22,4}^{224}]_{ij}=
{\pi\over 6\tau}(1-\delta_{i0})(1-\delta_{j0}).
\label{solution0}
\ee
These values were used as the initial guess, and starting from a low density the density was 
gradually increased to the desired values.

We consider the model with $\Delta=0.1$ and $\lambda=1$ at three different values of the
stickiness parameter $\tau=0.5,\;0.1$ and $0.04$ and at two different values of the number
density, $\rho=0.4,\;0.8$. 
In Fig. \ref{f1}, we show the fractions of i-times bonded particles $x_i$ as functions of density. As expected, increasing density or decreasing $\tau$ enhances bonding: the fraction of free particles $x_0$ decreases monotonically, whereas the fraction of four-times bonded particles $x_4$ increases. At the same time, the fractions of particles with an intermediate number of bonds, $x_1$, $x_2$ and $x_3$, initially increase with increasing density and, after reaching their maximum values, decrease and approach zero. Moreover, the maximum of
$x_{i-1}$ occurs at lower densities than that of $x_i$ ($i=2,3$). For lower values of $\tau$ positions of these maxima shift to the lower densities. This behavior demonstrates that the formation of the network of bonded particles proceeds from small clusters to a percolating, system-spanning network.

Figure \ref{f2} shows the the radial distribution function (RDF) 
${\tilde g}^{000}_{00}(r)$ and projections ${\tilde h}^{22l}_{22}(r)$. In figure \ref{f3} 
we display the corresponding RDF $g^{000(sw)}_{00}(r)$ and the projection $h^{224(sw)}_{22}(r)$ for the original square-well model.
In addition, these figures also include results obtained using the single-density approach \cite{cummings1986analytic}. Note that some expressions in Ref. \cite{cummings1986analytic} contain misprints; the corrected versions are provided in the Appendix.
At the high value of $\tau=0.5$, the predictions of the present theory closely match those of Cummings and Blum \cite{cummings1986analytic}, with noticeable differences observed only for
${\tilde h}^{224}_{22}(r)$ and $h^{224(sw)}_{22}(r)$ (figures \ref{f2} and \ref{f3}, left
panels). 
At this value of $\tau$, only about $4\%$ of the particles at $\rho=0.4$ and about $30\%$ at
$\rho=0.8$ 

\onecolumngrid
\noindent
\begin{figure}[htbp]
	\begin{center}
		\includegraphics[height=3.9823cm,clip]{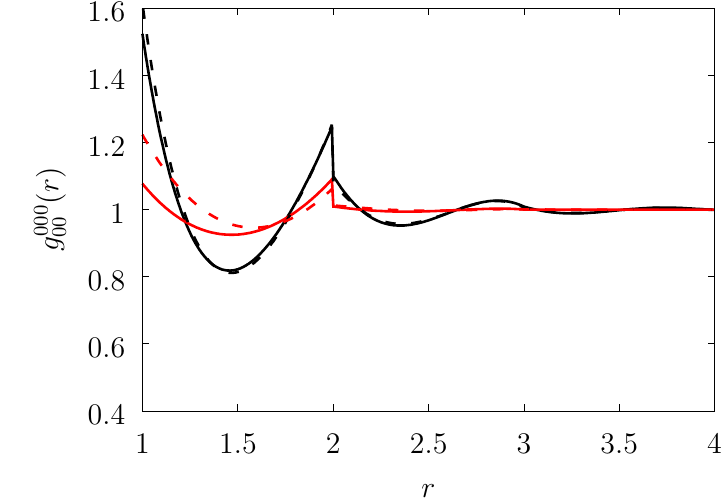}
		\includegraphics[height=3.9823cm,clip]{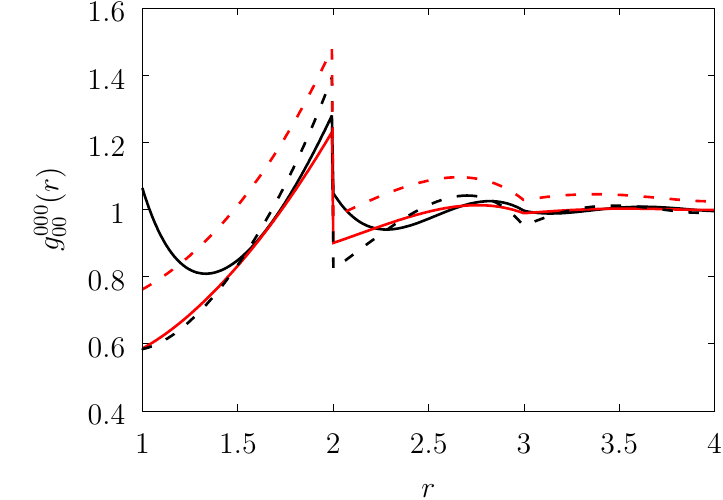}
		\includegraphics[height=3.9823cm,clip]{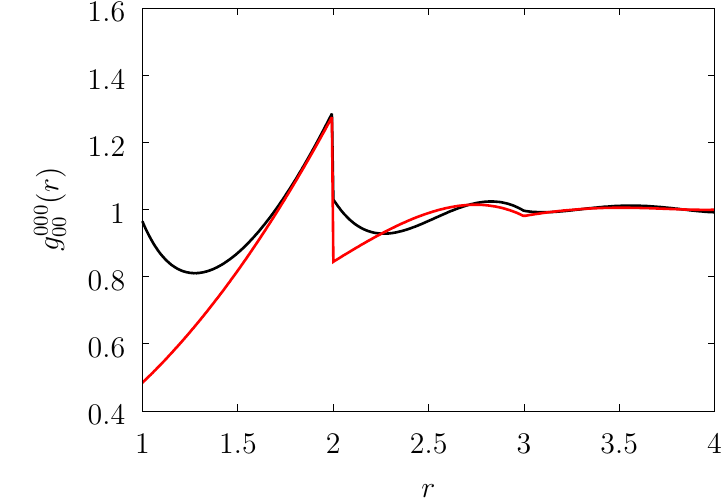}\\
		\includegraphics[height=3.9823cm,clip]{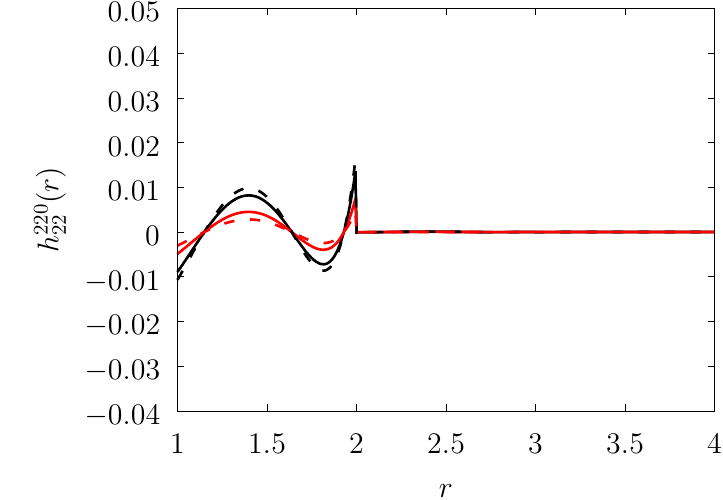}
		\includegraphics[height=3.9823cm,clip]{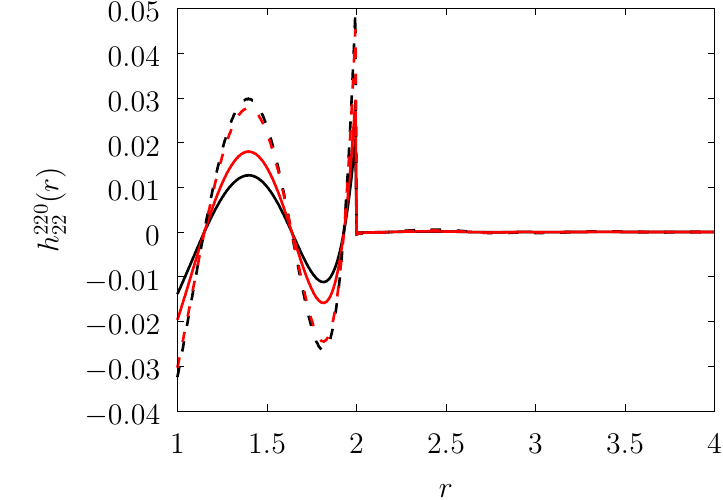}
		\includegraphics[height=3.9823cm,clip]{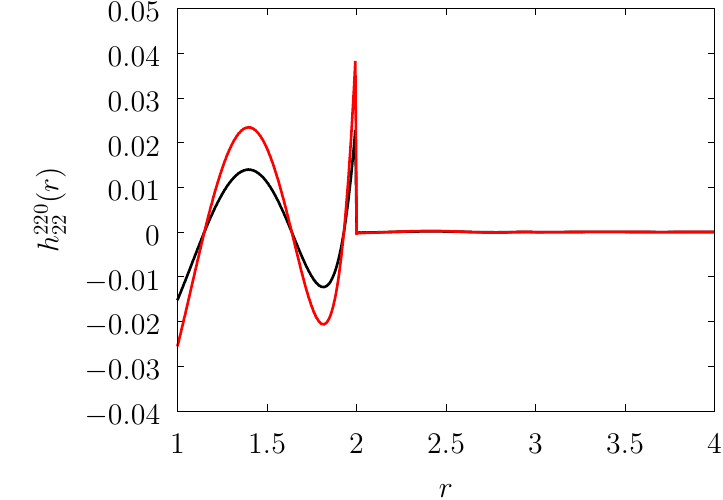}\\
		\includegraphics[height=3.9823cm,clip]{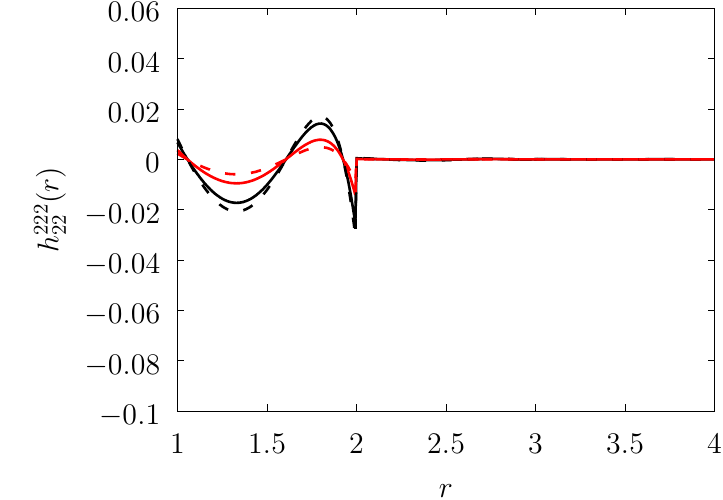}
		\includegraphics[height=3.9823cm,clip]{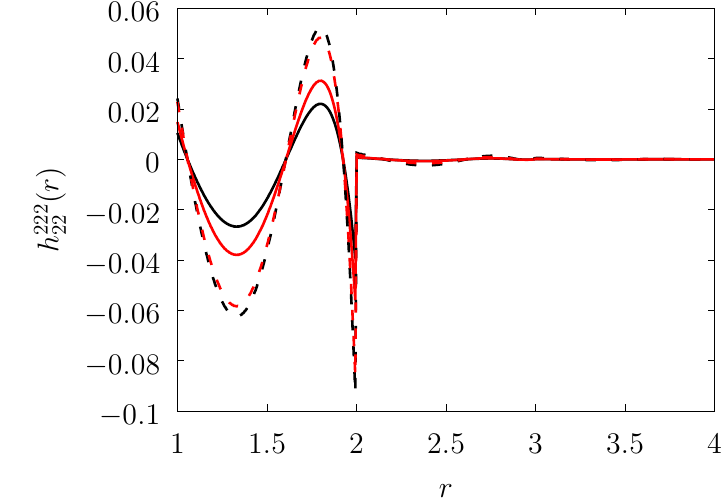}
		\includegraphics[height=3.9823cm,clip]{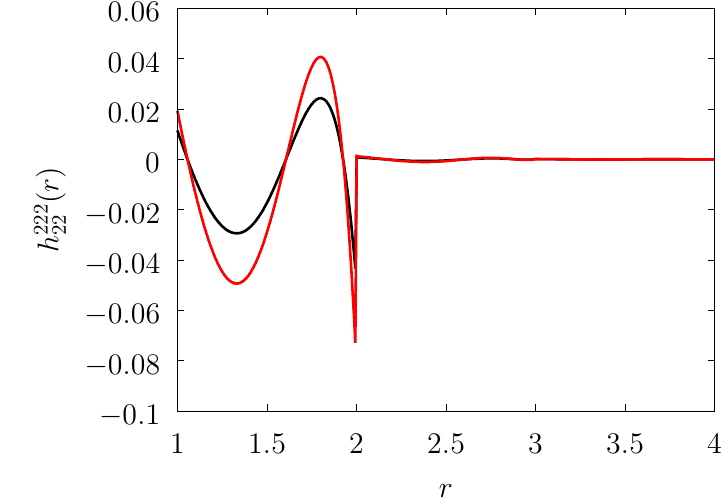}\\
		\includegraphics[height=3.9823cm,clip]{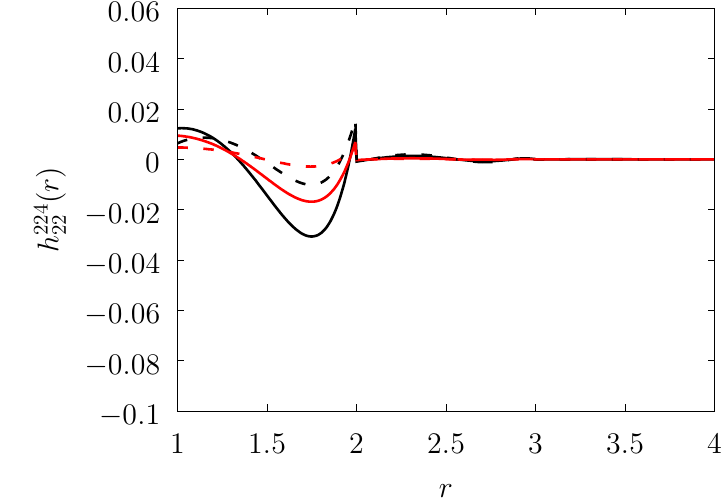}
		\includegraphics[height=3.9823cm,clip]{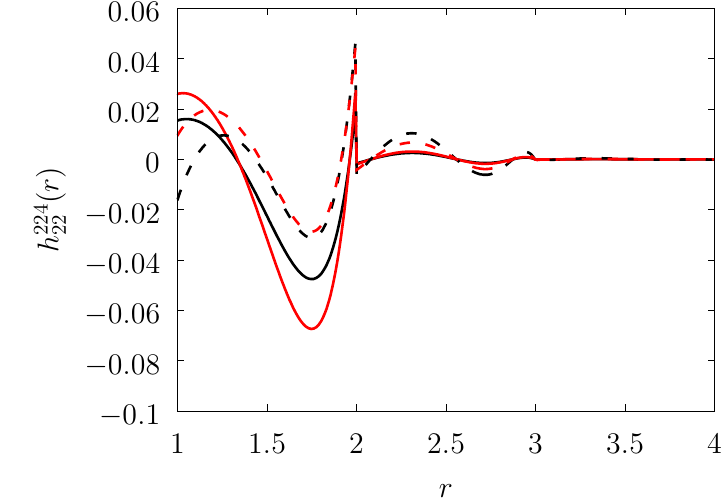}
		\includegraphics[height=3.9823cm,clip]{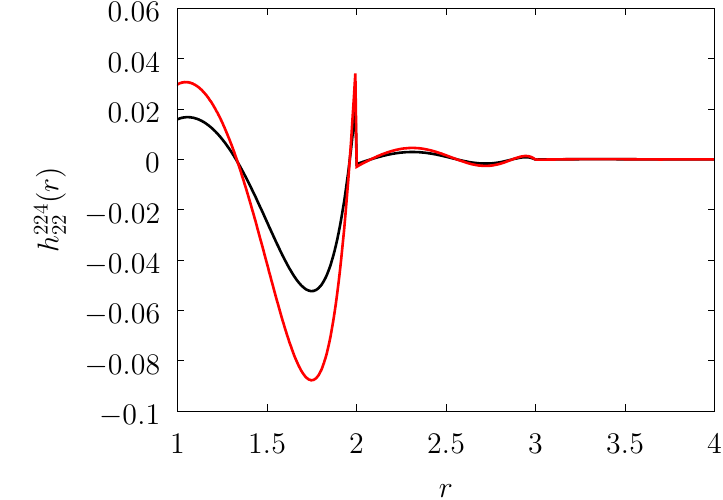}
	\end{center}
	\caption{ The radial distribution function $g^{000}_{00}(r)$ and projections 
		$h^{22l}_{22}(r)$ for $\tau=0.5$ (left column of panels), 
		$\tau=0.1$ (middle column of panels) and $\tau=0.04$ (right column of panels). 
		Here $\rho=0.4$ (red lines), $\rho=0.8$ (black lines), solid lines represent 
		results of the present mult-idensity theory and dashed lines represent results of 
		the single-density theory \cite{cummings1986analytic}.}
	\label{f2}
\end{figure}
\twocolumngrid
\noindent
are four-times bonded (see Fig. \ref{f1}, left panel). Upon decreasing 
$\tau$ to 0.1, the differences between our predictions and those of Cummings and Blum \cite{cummings1986analytic} become noticeable for all harmonic expansion coefficients, with the coefficients obtained from their theory exhibiting more pronounced structure (Figs. \ref{f2} and \ref{f3}, middle panels). At this intermediate stickiness, the degree of bonding is moderate, with $x_4\approx 0.37$ at $\rho=0.4$ and $x_4\approx 0.66$ at $\rho=0.8$.
Further decreasing $\tau$ to 0.04 results in a highly bonded system, with $x_4\approx 0.6$ at
$\rho=0.4$ and $x_4\approx 0.8$ at $\rho=0.8$. At this value of $\tau$, the approach of Cummings and Blum \cite{cummings1986analytic} does not converge, and only the predictions of the present theory are shown in Figs. \ref{f2} and \ref{f3} (right panels).
It is interesting to note that changing $\tau$ from a high value (0.5) to an intermediate one (0.1) produces noticeable changes in the RDF and harmonic expansion coefficients calculated using our theory. However, further decreasing $\tau$ to 0.04 results in only minor changes. This behavior reflects a key feature of the multidensity approach employed here \cite{kalyuzhnyi1993effects}: the maximum number of bonds per particle is limited and cannot exceed four, which is a reasonable assumption for the present model. At $\tau=0.1$, the degree of bonding is already relatively high, so further decreases in $\tau$ do not significantly alter the system’s structure.

\section{Conclusions}

In this paper, we formulate a multidensity version of the Ornstein–Zernike (OZ) equation, together with the corresponding Percus–Yevick (PY) closure relations, for a model of hard spheres with anisotropic surface adhesion of tetrahedral quadrupolar-like symmetry. We propose an analytical solution of this equation using the invariant expansion method combined with Baxter’s factorization technique. Numerical calculations are performed using both our solution of the multidensity OZ equation and the previously developed solution of the single-density molecular OZ equation \cite{cummings1986analytic}. The resulting pair correlation functions obtained from the single- and multidensity approaches are compared and analyzed. For low stickiness, the differences between the predictions of the two theories are small. As the stickiness increases, these differences become more pronounced, and beyond a certain value, the single-density approach no longer converges. In a subsequent publication, we plan to compare the predictions of both theories with exact computer simulation results and to extend our solution to a model including an additional point dipole, using a multidensity version of the mixed MSA/PY approximation. These studies are currently in progress and will be reported in due course.

\section{Acknowledgement}

Y.V.K. acknowledges financial support through the MSCA4Ukraine project (ID: 101101923), funded by the European Union. The hospitality of Vanderbilt University and Heriot-Watt University, where part of this work was carried out, is gratefully acknowledged.

\onecolumngrid

\begin{figure}[htbp]
	\begin{center}
		\includegraphics[height=4.cm,clip]{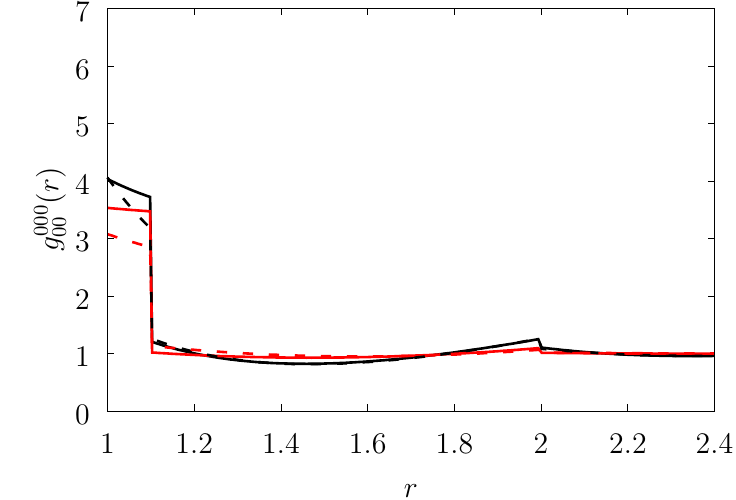}
		\includegraphics[height=4.cm,clip]{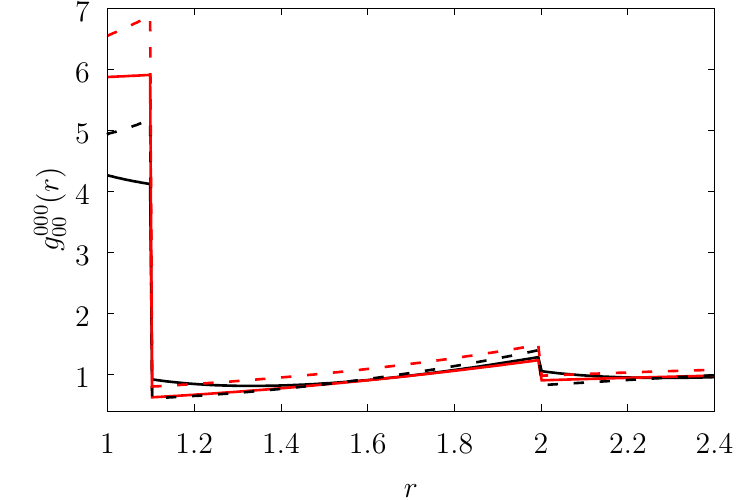}
		\includegraphics[height=4.cm,clip]{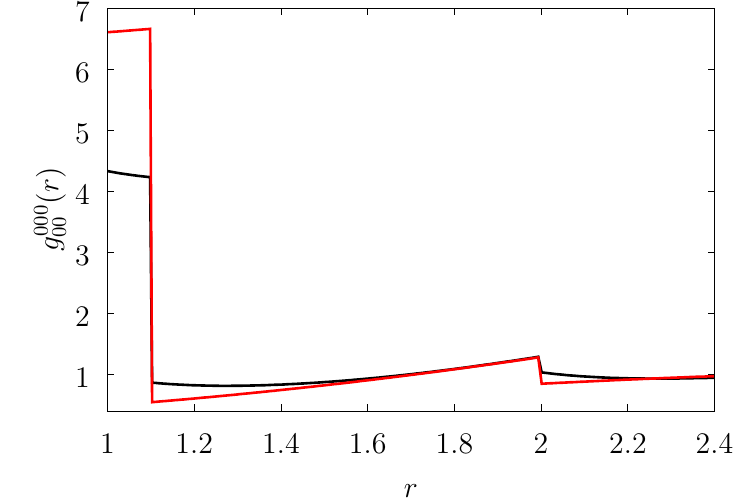}\\
		\includegraphics[height=4.cm,clip]{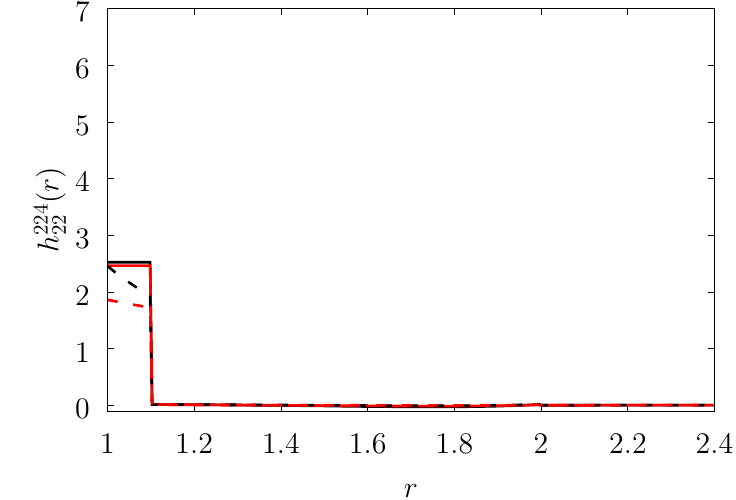}
		\includegraphics[height=4.cm,clip]{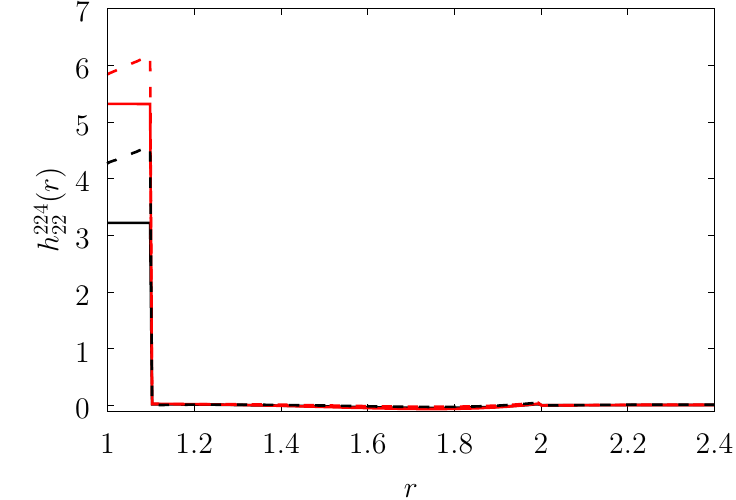}
		\includegraphics[height=4.cm,clip]{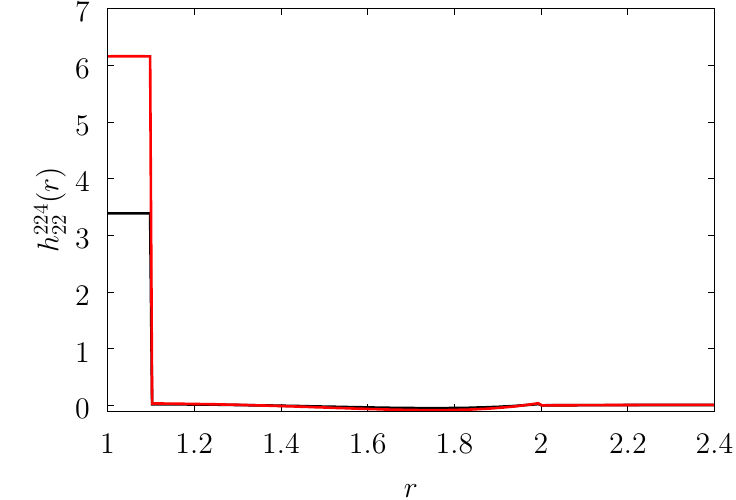}
	\end{center}
	\caption{ The radial distribution function $g^{000}_{00}(r)$ and projections 
		$h^{22l}_{22}(r)$ for $\tau=0.5$ (left column of panels), 
		$\tau=0.1$ (middle column of panels) and $\tau=0.04$ (right column of panels). 
		Here $\rho=0.4$ (red lines), $\rho=0.8$ (black lines), solid lines represent 
		results of the present mult-idensity theory and dashed lines represent results of 
		the single-density theory \cite{cummings1986analytic}.}
	\label{f3}
\end{figure}



\onecolumngrid

\appendix*
\section{Corrections of the misprints in \cite{cummings1986analytic}}

In revisiting the earlier work of Cummings and Blum \cite{cummings1986analytic}
the following corrections were found.

Eq. (1.5) of \cite{cummings1986analytic} should be:
\be
e(12)=
\left\{\begin{array}{ccccc}
	&0,&&r&<\sigma \\
	&1&+A\Phi_{00}^{000}(\omega_1\omega_2\omega_r)
+B\left[\Phi_{22}^{224}(\omega_1\omega_2\omega_r)+\Phi_{2{\underline 2}}^{224}(\omega_1\omega_2\omega_r)
+\Phi_{{\underline 2}2}^{224}(\omega_1\omega_2\omega_r)+\Phi_{{\underline 2}{\underline 2}}^{224}(\omega_1\omega_2\omega_r)
\right]
,\;
	 \sigma\leq &r& <\sigma+\omega\\
	 &1,&\;&r&\ge\sigma+\omega
\end{array} \right.,
\label{Aconical}
\ee
eq. (2.2) of \cite{cummings1986analytic} should be:
\be
H^{mm}_{\mu\mu;\chi}(k)=\int_0^\infty\;dr
\left[e^{ikr}J^{mm}_{\mu\mu;\chi}(r)
+e^{-ikr}J^{mm}_{\mu\mu;\chi}(r)\right]=
2\int_0^\infty\;dr\cos{(kr)}J_{\mu\mu;\chi}^{mm}(r)
\ee
eqs. (2.25), (2.26a) and (2.26b) of \cite{cummings1986analytic} should be:
\be
\beta_{\chi,0}=
(-)^\chi\left[\begin{pmatrix}
2 & 2 & 0 \\ 
\chi &  \underline{\chi} &0 \\
\end{pmatrix}
b^{220}_{22,0}\right.
\left.
-{1\over 2}
\begin{pmatrix}
	2 & 2 & 2 \\ 
	\chi &  \underline{\chi} &0 \\
\end{pmatrix}
b^{222}_{22,0}
+{3\over 8}
\begin{pmatrix}
	2 & 2 & 4 \\ 
	\chi &  \underline{\chi} &0 \\
\end{pmatrix}
b^{224}_{22,0}\right],
\ee
\be
\beta_{\chi,2}=(-)^\chi\left[
{3\over 2}
\begin{pmatrix}
	2 & 2 & 2 \\ 
	\chi &  \underline{\chi} &0 \\
\end{pmatrix}
b^{222}_{22,2}
-{15\over 4}
\begin{pmatrix}
	2 & 2 & 4 \\ 
	\chi &  \underline{\chi} &0 \\
\end{pmatrix}
b^{224}_{22,2}\right],
\ee
\be
\beta_{\chi,4}=(-)^\chi
{35\over 8}
\begin{pmatrix}
	2 & 2 & 4 \\ 
	\chi &  \underline{\chi} &0 \\
\end{pmatrix}
b^{224}_{22,4},
\ee
eqs. (2.37a) and (2.37b) of \cite{cummings1986analytic} should be:
\[
\int_0^1\;dr_1S_{22;\chi}^{22}(r_1)=-(1-2\rho\lambda_{2\chi})
\sum_{m=1}^4{1\over (m-1)!(m+1)}q_{\chi,m}^{22}
\]
\be
-2\rho\sum_{m=1}^4\sum_{n=1}^4
{1\over m!n!}\left[{1\over (m+n+2)(n+1)}-{1\over m+2}-
{1\over (n+2)(n+1)}+{1\over 2}\right]q_{\chi,m}^{22}q_{\chi,n}^{22}+
\lambda_{2\chi}-\rho\lambda_{2\chi}^2,
\label{AS1}
\ee
\[
\int_0^1\;dr_1r_1^2S_{22;\chi}^{22}(r_1)=-\sum_{m=1}^4
{1\over 3(m-1)!(m+3)}q^{22}_{\chi,m}
\]
\[
-2\rho\sum_{m=1}^4\sum_{n=1}^4
{1\over m!n!}\left[{2\over (m+n+4)(n+3)(n+2)(n+1)}-{1\over 3(m+4)}-
	{2\over (n+4)(n+3)(n+2)(n+1)}+{1\over 12}\right]q^{22}_{\chi,m}
	q^{22}_{\chi,n}
\]
\be
+{1\over 3}\lambda_{2\chi}-2\rho\lambda_{2\chi}\sum_{m=1}^4{1\over m!}
\left[{1\over 3(m+4)}+{2\over (m+4)(m+3)(m+2)(m+1)}-{1\over 6}\right]
	q_{\chi,m}^{22}-{\rho\over 6}\lambda_{2\chi}^2,
\label{AS2}
\ee
eq. (2.41a) of \cite{cummings1986analytic} should be:
\[
{\hat c}^{000}_{00}(r)={\hat f}^{000}_{00}(r)\left[{\hat g}_{00}^{000}(r)-
{\hat c}_{00}^{000}(r)\right]+
4{\hat f}^{220}_{22}(r)\left[{\hat g}_{22}^{220}(r)-
{\hat c}_{22}^{220}(r)\right]
\]
\be
+{4\over 5}{\hat f}^{222}_{22}(r)\left[{\hat g}_{22}^{222}(r)-
{\hat c}_{22}^{222}(r)\right]+
{4\over 9}{\hat f}^{224}_{22}(r)\left[{\hat g}_{22}^{224}(r)-
{\hat c}_{22}^{224}(r)\right]
\ee

\twocolumngrid
\noindent

\bibliography{PTCH2O}

\end{document}